\documentclass[showkeys,twocolumn,nofootinbib]{revtex4}
\usepackage{subscript}
\usepackage{natbib}
\usepackage{amssymb}
\usepackage{color}
\usepackage{amsfonts}
\usepackage{textcomp}
\newcommand{\ped}[1]{\ensuremath{_{\rm #1}}}
\newcommand{\apex}[1]{\ensuremath{^{\rm #1}}}
\definecolor{blue}{rgb}{0,0,0}
\usepackage{float}
\usepackage[caption = false]{subfig}
\usepackage{graphicx}

\begin{document}

\title{Carrier mobility and scattering lifetime in electric double-layer gated few-layer graphene}

\author{E. Piatti}
\author{S. Galasso}
\author{M. Tortello}
\author{J. R. Nair}
\author{C. Gerbaldi}

\affiliation{Dipartimento di Scienza Applicata e Tecnologia, Politecnico di Torino, 10129 Torino, Italy}

\author{M. Bruna$^+$}
\author{S. Borini$^+$}

\affiliation{Istituto Nazionale di Ricerca Metrologica (INRIM), 10135 Torino, Italy}

\author{D. Daghero}
\author{R. S. Gonnelli}
\email{renato.gonnelli@polito.it}

\affiliation{Dipartimento di Scienza Applicata e Tecnologia, Politecnico di Torino, 10129 Torino, Italy}


\begin{abstract}
We fabricate electric double-layer field-effect transistor (EDL-FET) devices on mechanically exfoliated few-layer graphene. We exploit the large capacitance of a polymeric electrolyte to study the transport properties of three, four and five-layer samples under a large induced surface charge density both above and below the glass transition temperature of the polymer. We find that the carrier mobility shows a strong asymmetry between the hole and electron doping regime. We then employ \emph{ab-initio} density functional theory (DFT) calculations to determine the average scattering lifetime from the experimental data. We explain its peculiar dependence on the carrier density in terms of the specific properties of the electrolyte we used in our experiments.
\end{abstract}


\maketitle

\section{Introduction}
{\let\thefootnote\relax\footnotetext{\textsuperscript{+}Current address: Nokia Technologies, CB30FA Cambridge, United Kingdom}}

Since its discovery in 2004 \cite{Novoselov04}, graphene has attracted a widespread interest from the scientific community due to its peculiar physical and electronic properties. Moreover, it has soon been discovered that graphene crystals having a different number of layers effectively behave as completely different electronic systems, approaching bulk graphite for a number of layers $ N\gtrsim 8$ \cite{FaiMak10}. While a lot of effort has been devoted to the characterization of single-layer graphene (SLG), few-layer graphene (FLG) has received a smaller attention \cite{CastroNeto09}. Furthermore, most of the investigations has been restricted to the relatively small carrier density range close to the Dirac point, where the relativistic properties of the graphene quasiparticle excitations diverge more strongly from those of standard semiconductors.

\begin{figure*}
\begin{center}
\includegraphics[keepaspectratio, width=0.7\textwidth]{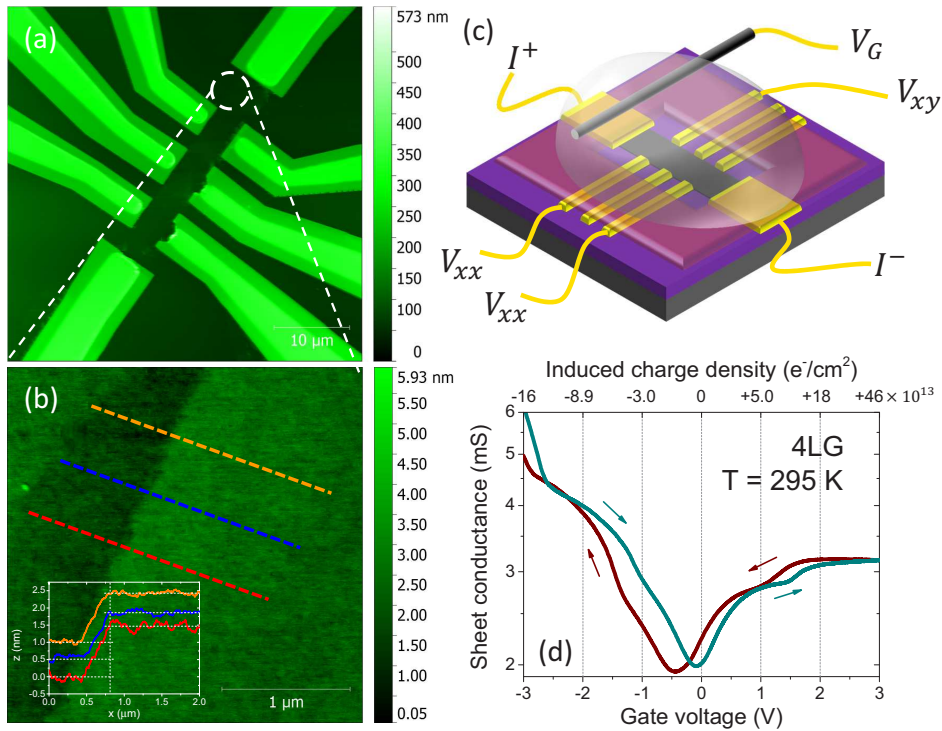}
\end{center}
\caption {
(a) AFM topographic image in contact mode of a 3LG device before PES drop-casting, and (b) zoom on its upper left corner. Inset shows the height signal step in correspondence to the graphene edge. While signs of electrical damage occurred during preliminary magnetoresistance measurements appear in the lower part of the device, the entire middle-to-upper regions are free from wrinkles and folds. The AFM-estimated thickness agrees well with the results obtained via optical characterization. (c) Sketch of a complete device after PES drop-casting. Gate voltage is applied between the platinum wire and the negative current contact. (d) Sheet conductance vs gate voltage transfer curves for a 4LG sample in the [-3V, +3V] range at room temperature. Top scale shows the corresponding induced charge density estimated from DSCC.} \label{figure:device}
\end{figure*}

The investigation of carrier-density ranges further away from the Dirac point requires a larger tunability of the Fermi level in a reproducible way. This can be achieved by the technique of electric double-layer (EDL) gating, which has been introduced in the field of solid-state physics since the second half of the 2000s. With respect to standard field-effect techniques, EDL gating is able to access values of carrier density previously unattainable in chemically undoped insulators and semiconductors \cite{UenoJPSJ14}. This strong increase in the tunability of the carrier density in a field effect transistor (FET) architecture is made possible by the strong increase in the gate capacitance. In turn this is obtained by substituting the solid gate oxide with a liquid or polymeric electrolyte (this is the reason why the technique is also referred to as liquid gating). Indeed, after the first results on strontium titanate \cite{Ueno08}, EDL gating has soon been employed on the prototypical 2D materials, first on SLG \cite{Efetov10} and soon after on double and three-layer graphene (2LG and 3LG respectively) \cite{YePNAS}. However, the properties of liquid-gated few-layer graphene of higher number of layers have only recently been explored in our earlier work of Ref. \cite{Gonnelli15}. In that paper we focused on the low-temperature transport properties of 3LG, 4LG and 5LG.

In the present work, we focus instead on how the electric transport properties of few-layer graphene are modified by EDL gating at higher temperatures, both immediately above and below the glass phase transition of our electrolyte. We find that our samples show a marked asymmetry in the response to electron and hole doping, i.e. to a positive or negative applied gate voltage. In particular, while the mobility is suppressed in our devices by both positive and negative applied gate voltages, the suppression in the electronic side is much more pronounced than in the holonic one. By combining our experimental findings with \emph{ab-initio} density functional theory (DFT) calculations, we determine how the average scattering lifetime in our devices is affected by the strong modulation of the carrier density. We discover that the asymmetry in the response is even more pronounced for the scattering lifetime than for the mobility. In addition a difference in the carrier-density dependence of the scattering lifetime emerges between 4LG and 5LG. Finally, we propose a qualitative explanation for these findings, in which we link the asymmetry in the transport properties of the material to the asymmetry in the physical dimensions of the active ions composing our electrolyte.

\section{Device fabrication}
We obtained the few-layer graphene samples via mechanical exfoliation of highly oriented pyrolitic graphite (HOPG) and transferred them onto standard SiO\textsubscript{2} on Si substrates. The number of layers of each flake was determined by optical contrast analysis \cite{Bruna09}; Raman spectroscopy was also employed to independently confirm this number and to determine the stacking sequence \cite{Lui11b, Ferrari_NNano2013, Ferrari2006, Cong2011}. All the samples that will be discussed in the following are 3LG, 4LG and 5LG presenting Bernal stacking. We realized the electrical contacts for the four-wire resistance measurements via the standard microfabrication techniques, i.e. photolithography, Cr/Au thermal evaporation and lift-off. Subsequently we patterned the devices into a Hall-bar shape by reactive ion etching. An additional windowing was finally realized on the device by photoresist in order to leave uncovered only a part of the contact pads (for wire bonding) and the gated channel over the flake. The photoresist mask was hard-baked at 145\textdegree C for 5 minutes to improve its electrochemical stability. Figure \ref{figure:device}a shows an AFM image of a complete device after it underwent a preliminary magnetoresistance characterization, while Figure \ref{figure:device}b shows a magnification of the upper left corner of the device to allow for easier thickness determination. The AFM characterization shows that the device is free from intrinsic defects such as wrinkles and folds. The flake thickness ($\approx$ 1.4 nm, corresponding to a 3LG sample \cite{Koh2011}) estimated at the edge (see inset to Figure \ref{figure:device}b) agrees well with the results obtained from the optical characterization.

After wire bonding we drop-casted the liquid precursor of the polymer electrolyte system (PES) onto the complete devices in the controlled atmosphere of a dry room in order to avoid contamination from water molecules and other chemicals that may affect the performance of the electrolyte. We employed the same Li-TFSI based PES used in our earlier experiments on gold \cite{Daghero12} and other noble metals \cite{Tortello13} due to its record amount of charge induction. Figure \ref{figure:device}c shows a sketch of a device, where the few-layer graphene flake acts as the working electrode of the electrochemical cell, while a Pt wire is immersed in the PES and acts as the counter electrode. The electrical connections for longitudinal (R\textsubscript{xx}) and Hall (R\textsubscript{xy}) resistance measurements are shown as well.

\section{Experimental results}
We performed gating and transport measurements in a Cryomech$^{\circledR}$ PT405 pulse-tube cryocooler. After PES drop-casting the devices were rapidly transferred inside the cryostat in order to minimize their exposure to air. We left the devices in vacuum at room temperature for a few hours to degas before performing the actual measurements. The gate voltage was then applied at room temperature both in form of ramps and step perturbations, and the response of the samples resistance and of the gate-current was recorded. Slow gate-voltage ramps were employed to assess the ambipolar nature of the field effect in all our samples. It means that we always detected the electron (hole) conduction in the channel upon application of a sufficiently large positive (negative) bias between the gate electrode and the sample. Step-like gate-voltage applications were instead used to allow for double-step chronocoulometry (DSCC) measurements able to determine the induced carrier density in the channel of our devices. A detailed analysis of this technique, the comparisons with the standard Hall effect measurements, and the assessment on the reproducibility and purely electrostatic nature of the gating technique can be found in our earlier work \cite{Gonnelli15}. The total sheet carrier density in the system $n_{2D}$ is determined by adding the induced carrier density to the pristine density measured by Hall effect.

Figure \ref{figure:device}d shows a typical sheet conductance response of a 4LG device both as a function of the applied gate voltage and of the induced carrier density in the $[-3V, +3V]$ voltage range. The small negative hysteresis between an increasing or decreasing gate-voltage ramp is typical of electrolyte-gated graphene devices and can be attributed to carrier transfer from the metal contacts due to capacitive gating \cite{WangNano2010}. This also accounts for the minimum in the conductance curves being shifted to negative gate voltages (indicating that the sample is n-doped). In fact Hall effect measurements performed immediately after drop-casting showed a native p-doping of the sample, as is to be expected in the case of traces of residual water adsorbed at the graphene surface. Two other features clearly emerge from these curves: first, both for electron and hole doping a sharp slope change arises for intermediate values of the applied gate voltage ($V_G\approx\pm1.5V$). This behavior has already been observed in 3LG \cite{YePNAS} and linked to the rise of interband scattering due to the crossing of a Van Hove singularity when the Fermi level crosses the split band T\textsubscript{2g}. Second, while the curves are symmetric for low values of the applied gate voltage, a strong asymmetry emerges for higher values. The electron-doped branch saturates at almost half the sheet conductance value with respect to the maximum one in the hole-doped branch, even though the induced carrier density is estimated to be significantly larger.

\begin{figure}
\begin{center}
\includegraphics[keepaspectratio, width=0.8\columnwidth]{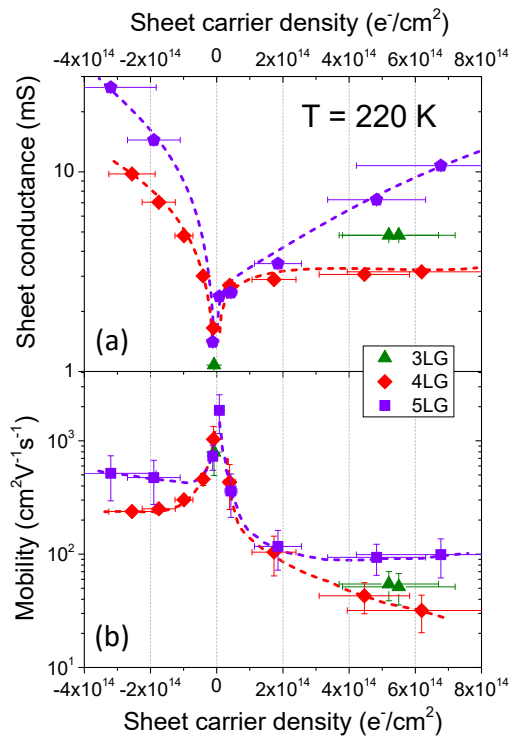}
\end{center}
\caption {
Electrical transport characterization of the device at  T = 220 K, below the glass transition of the electrolyte. Sheet conductance (a) and mobility (b) vs carrier density $n_{2D}$ for 3LG (green), 4LG (red) and 5LG (violet) devices. The strongly asymmetric response of the system to electron and hole doping is readily distinguishable. Dashed lines act as guides to the eye.} \label{figure:experimental}
\end{figure}

We then studied in more detail the transport properties of our samples for few selected values of the applied gate voltage. For each of these values, the voltage was applied at room temperature in order to measure the DSCC response, and the sample was subsequently cooled down to the base temperature of the cryogenic system. An analysis of the transport properties of FLG at low temperature is beyond the scope of this paper and has been detailed elsewhere \cite{Gonnelli15}. Here, we focus on the regime around 220 K, i.e. where the sample is still relatively close to room temperature but the PES has already undergone its glass transition. This ensures that the ion movement inside the electrolyte is completely frozen, thus minimizing the fluctuations in the charge density accumulated at the sample surface.

Figure \ref{figure:experimental}a shows the dependence of the sheet conductance on the sheet carrier density $n_{2D}$ at 220 K, as determined by DSCC, for both the 4LG and 5LG samples. Data from a 3LG sample are shown for comparison, but will not be discussed due to the small number of experimental points. In the case of 4LG, the behavior at 220K closely resembles the response to the gate sweep at room temperature. In particular, the sheet conductance on the electron side shows clear signs of saturation already at relatively small values of carrier density, while on the hole side an incresing trend is observed in the entire density range. As a consequence the maximum conductance value on the hole branch is almost three times larger than the one on the electron branch. Interestingly, the asymmetry is much less pronounced in the case of 5LG. In this case no sign of saturation is evident also on the electron side and the maximum conductance in the hole branch is only about twice as big as that in the electron one.

This dependence of the difference in the asymmetry on the number of layers is present also in the dependence of the mobility $\mu =  \sigma_{2D}/(en_{2D})$ on $n_{2D}$, as shown in Figure \ref{figure:experimental}b. The mobility decreases at the increase of the carrier density for both electron and hole doping, independently of the number of layers. However, in the case of 5LG the mobility rapidly reaches saturation. In the case of 4LG instead, the mobility decrease is significantly more pronounced, and signs of saturation are present only in the hole branch.

\section{Discussion}
As we have seen, three peculiar behaviors emerge from the carrier density dependence of the mobility. First, $\mu$ decreases with increasing $n_{2D}$ for both electron and hole doping. Second, this dependence is strongly asymmetric in the sign of the charge carriers. Third, this asymmetry is less pronounced in the case of 5LG with respect to 4LG.

We first consider the decrease in mobility. In SLG the mobility is expected to decrease moving away from the Dirac point. However, in 2LG and 3LG it has been shown that the different density of states leads to an increasing behavior, albeit in a smaller range of carrier densities than the one considered in this paper \cite{ZhuPRB09}. On the other hand, it is known that in two-dimensional electron gases (2DEGs) in general, and in graphene in particular, a re-entrant mobility at very high carrier density is associated to a crossover from a dominant scattering by screened Coulomb charges to a scattering dominated by surface roughness (for 2DEGs in semiconductors \cite{DasSarmaPRB14}) or short-range scatterers (in graphene \cite{DasSarmaRMP11}).

\begin{figure}[t]
\begin{center}
\includegraphics[keepaspectratio, width=0.8\columnwidth]{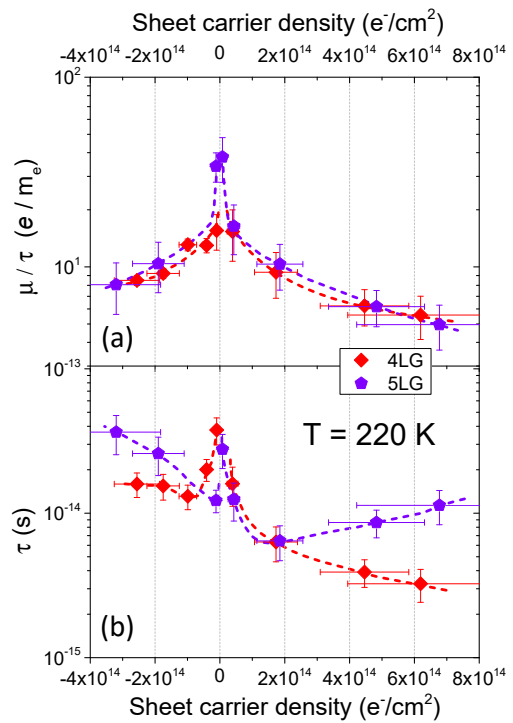}
\end{center}
\caption{
(a) Dependence of the mobility-over-scattering lifetime ratio, $\mu/\tau$, as function of the carrier density $n_{2D}$ as estimated from DSCC measurements and DFT calculations, in units of the elementary charge divided by the electron mass. The behavior is symmetric with respect to electron and hole doping and strongly depends on the number of layers only in close vicinity to the Dirac point. (b) Average scattering lifetime $\tau$ vs carrier density $n_{2D}$ calculated by dividing the experimentally measured mobility by $\mu/\tau$. The asymmetry of $\tau$ between electron and hole doping is enhanced with respect to that of the experimental mobility, and a qualitative difference in the response of 4LG and 5LG is also readily apparent. Results are reported for 4LG (red) and 5LG (violet). As in Figure\ref{figure:experimental}, dashed lines are guides to the eye.} \label{figure:discussion}
\end{figure}

\textcolor{blue}{In order to disentangle the contributions to the mobility due to the changes in the DOS from those arising from changes in the scattering rates, we combine our experimental results with \emph{ab-initio} DFT calculations \cite{Jones89,DFTbooks}. The bandstructure of 4LG and 5LG under strong doping conditions is calculated by using the all-electron full-potential linearised augmented-plane wave method as implemented in the ELK code \cite{ELKsite}. We model the FLG by building a 3D supercell with a value of the \emph{c}-axis lattice constant larger than 20 $\rm \AA$ so that, in order to avoid interactions, the periodic images of the structures are at least 10 $\rm \AA$ apart. Independently of the number of layers the layer to layer distance is taken equal to 3.35 $\rm \AA$ while the in-plane lattice constant is fixed to a = 2.46 $\rm \AA$. As a first approximation we simulate the effect of the EDL gating by introducing extra charge carriers into the system together with a compensating homogeneous background charge (Jellium model). A more complete model, that takes into account the intense electric field at the EDL/graphene interface \cite{Brumme14, Brumme15} goes beyond the scope of this paper. In order to describe the exchange and correlation terms we adopted the local
density approximation (LDA) and the hexagonal Brillouin zone was sampled with a 28x28x1 mesh of k-points. A total energy tolerance of $10^{-8}$ Hartree was used for obtaining the convergence of the self-consistent field calculations.}

\textcolor{blue}{It is well known that the effective mass $m^*$ is not a good quantity for the description of the massless relativistic quasiparticles close to the Dirac point of graphene. However, an expression for the mobility-to-scattering lifetime ratio can be obtained knowing the theoretical bandstructure and the experimental value of $n_{2D}$ according to \cite{Cappelluti09}:}

\begin{equation}
\frac{\mu}{\tau} = \frac{1}{\tau} \frac{\sigma_{2D}}{en_{2D}} =
\frac{e}{n_{2D}} \sum_i (\frac{N\ped{k\ped{i}} N\ped{s}}{4} v\ped{F\ped{i}}\apex2 N\ped{i})
\label{sigma_B}
\end{equation}

where $N\ped{k\ped{i}}$ and $N\ped{s}$ are the valley and spin degeneracies, respectively, $v\ped{F\ped{i}}$ is the Fermi velocity and $N\ped{i}$ the DOS at the Fermi level of the i-th band.

Unsurprisingly, the symmetry of the bandstructure of FLG above and below the Dirac point is reflected on the symmetric dependence of the ratio $\mu/\tau$ on the carrier density (Figure \ref{figure:discussion}a). Thus, the strong asymmetry in the doping dependence of the mobility cannot be attributed to features in the bandstructure. However, the clear reduction of $\mu/\tau$ moving away from the Dirac point can be partly responsible for the similarly decreasing behavior of the mobility.

We finally extract the average scattering lifetime $\tau$ by dividing the measured mobility by the DFT-determined ratio $\mu/\tau$ (Figure \ref{figure:discussion}b). In this case, the asymmetries in the electron- and hole-doping branches are magnified, as well as the difference in the behaviour between 4LG and 5LG. In 4LG, the scattering lifetime monotonically decreases across the entire electron branch, while rapidly saturates after an initial decrease in the hole branch. Instead, in 5LG $\tau$ shows a nonmonotonic behavior in the electron branch, decreasing rapidly for small values of carrier density and increasing for larger ones, while in the hole branch it appears to be monotonically increasing.

This very peculiar behavior cannot be simply associated with modifications in the DOS, nor to a pure crossover from screened Coulomb scattering to short-range scattering. Instead, an intrinsically surface-sensitive, asymmetric scattering mechanism is required to account for our findings. In the following, we propose a possible explanation that is directly related to the physical implementation of the EDL technique.

\textcolor{blue}{As we mentioned earlier, when the gate voltage is applied, the solvated ions inside the polymeric matrix densely accumulate in close vicinity (\ensuremath{\lesssim} 1 nm) to the graphene surface, thus building a nanoscale capacitor at the electrolyte/graphene interface. Since it has been shown \cite{ZhangNatPhys09} that charged impurities in close proximity to the graphene surface create scattering centers for Dirac quasiparticles, we can assume that the ions constituting the EDL itself may behave in the same way.}

As long as the electric transport is dominated by screened Coulomb scattering, the average $\tau$ of the system is determined by a competition between two contributions. On one hand, the increasing number of charge carriers introduced by gating improves the screening capabilities of the 2DEG and would enhance $\tau$ and $\mu$. On the other hand, the concomitant increasing number of defects introduced by the presence of charged ions at the graphene surface would decrease $\tau$ and $\mu$. The reduced suppression of the mobility with increasing carrier density in 5LG would then be a direct consequence of its reduced surface to volume ratio with respect to 4LG.

The strong asymmetry of the conductance, the mobility, and the scattering lifetime between electron and hole doping in both 4LG and 5LG can instead be related to the asymmetry in the size of the ions composing the Li-TFSI salt. In fact, the standard ionic liquids used in EDL gating literature (e.g. DEME-TFSI), have cations and anions that are both made by complex molecules of comparable size \cite{UenoJPSJ14}. In the case of Li-TFSI instead, the Li\textsuperscript{+} cation (inducing electron doping) is significantly smaller than its anion counterpart (inducing hole doping). This characteristic can allow it to pack at the electrode surface both with higher density and, more importantly, in closer proximity. This in turn enhances its charge-induction capabilities with respect to the anion, as demonstrated by our DSCC measurements on both graphene \cite{Gonnelli15} and noble metals \cite{Daghero12}. On the other hand this fact most likely makes its efficiency as a scattering center sensibly larger than that of the bigger TFSI anion.

\section{Conclusions}
In conclusion, we characterized the electric transport properties of 4LG and 5LG field-effect devices in the very high carrier-density regime achievable by EDL gating, both above and below the freezing point of the electrolyte. We detected a peculiar asymmetry in the response to hole and electron doping in both the conductance and the mobility of the samples. By combining our experimental results with \emph{ab-initio} DFT calculations, we extracted the dependence of the average scattering lifetime on both the number of layers and the carrier density. We have interpreted the resulting trends as primarily due to the microscopic properties of our EDL gating technique, i.e. the capability of the ions composing the EDL to act as extra charged scattering centers for carriers at the surface of the device. In particular, the strong asymmetry between electron and hole doping has been associated with the difference in size between cations and anions in the solvated salt Li-TFSI. These results are relevant in the framework of understanding the non trivial modifications the EDL technique causes at the surface of graphene and other two-dimensional materials.

\section*{Acknowledgments}
The authors acknowledge the support from the EU Graphene Flagship and Graphene@PoliTo. We are grateful to E. Cappelluti for tight-binding calculations we used to check the validity of DFT results.

\end{document}